\documentclass[preprint,12pt,3p]{elsarticle}
\usepackage[utf8]{inputenc} 
\usepackage[T1]{fontenc}    
\usepackage{hyperref}       
\usepackage{url}            
\usepackage{booktabs}       
\usepackage{amsfonts}       
\usepackage{nicefrac}       
\usepackage{microtype}      
\usepackage{lipsum}
\usepackage{amsmath,amssymb,graphicx}
\usepackage{float}
\usepackage{siunitx}
\usepackage{array}
\usepackage{lmodern}
\usepackage{amssymb}
\usepackage{xcolor}
\usepackage{multicol}
\usepackage{multirow}
\usepackage{orcidlink}
\usepackage{amsmath,amssymb,exscale}
\usepackage{amsmath}
\usepackage{amsmath,amssymb,exscale}
\usepackage[final]{pdfpages}
\usepackage[bottom]{footmisc}
\journal{Elsevier}\begin{document}
\begin{frontmatter}
\title{Poly-dodecahedrane: A new allotrope of carbon}
\author[Lorestan University]{Siavash Hasanvandi\corref{author}\,\orcidlink{0000-0002-8876-1271}}
\author[Lorestan University]{Elham Neisi\corref{author}\,\orcidlink{0009-0002-0794-3725}}
\author[IFPI]{José M. De Sousa\corref{author}\,\orcidlink{0000-0002-3941-2382}}
\cortext[author]{Corresponding authors: s.hasanvandi$@$gmail.com.com (Siavash Hasanvandi), elham.neisi$@$yahoo.com (Elham Neisi) and josemoreiradesousa$@$ifpi.edu.br (José M. De Sousa)}

\affiliation[Lorestan University]{organization={Department of Organic Chemistry, Faculty of Chemistry, Lorestan University},
            addressline={Khorramabad}, 
            city={Lorestan},
        country={Iran.}}
        
\affiliation[IFPI]{organization={Instituto Federal de Educa\c c\~ao, Ci\^encia e Tecnologia do Piau\'i -- IFPI},
            addressline={Primavera}, 
            city={São Raimundo Nonato},
            postcode={64770-000}, 
            state={Piauí},
            country={Brazil.}}

\begin{abstract}
Carbon is the most important chemical element and the theoretical study of its new allotropes can be of great interest.  In this study, regular dodecahedron (dodecahedrane) oligomers $(n = 1, 3, 5, 7, 9, 11, 13)$ by extending the dodecahedrane units in 3-dimensions were designed. Then, a theoretical study was conducted on their structures and electronic properties as potential new carbon allotrope. The cohesive energy ($Ecoh$) and $\Delta G$, were calculated. Experimental observations indicate that the $Ecoh$ rises as the number of dodecahedrane units increases, whereas the Gibbs free energy change $\Delta G$ decreases with an increase in the number of dodecahedrane units. The HOMO-LUMO energy gap ($Eg$) values, which represent electronic properties, decrease with increasing number of dodecahedrane units. Density functional theory (DFT) calculations of the novel carbon allotropes polydodecahedrane nanostructures have unveiled a previously unobserved symmetry, indicating intrinsic metallic behavior. The symmetrical distribution of partial charges was found in molecular electrostatic potential ($MEP$) diagrams for all oligomers, showing a tendency of the structures to maintain a symmetrical structural order as the number of monomer units increases. In addition, natural bond orbital ($NBO$) analysis of 13-units oligomer as largest designed structure reveals near-$sp^{3}$ hybridization for different carbons. Based on the calculated results, the structures have a tendency to extend in $3$-dimensions and form a covalent network of poly-dodecahedrane with a unique structure consisting of interconnected cyclopentane rings. The results show that this exclusive configuration exhibits theoretical stability and suggests the potential for poly-dodecahedrane to be regarded as a novel carbon allotrope.
\end{abstract}


\end{frontmatter}
\section{Introduction}
\label{INT}

Materials science is a dynamic and interdisciplinary field of research that is instrumental in the discovery of novel nanostructures with experimental and theoretical underpinnings \cite{shackelford2000materials,steinbach2009phase,mittemeijer2010fundamentals}. This dynamic discipline delves into the properties, behavior, and applications of a vast array of materials, ranging from traditional polymers to cutting-edge nanomaterials and composites \cite{zhu2018molecular,meyer2006nanotechnology}.

In this great area, the proposal and investigation of new allotropes (new archetypical lowdimensional carbon forms - 0D fullerenes, 1D nanotubes, and 2D graphene) derived from carbon was revolutionized by the discovery and synthesis of fullerenes \cite{kroto1985c60,kratschmer1990solid}, the subsequent identification of nanotubes \cite{iijima1991helical} and the isolation of a single layer of graphite (called graphene) \cite{novoselov2004electric}. These discoveries revealed that carbon materials-based nanostructures in condensed matter, formally found in graphite and diamond \cite{terrones2010graphene,haines2001synthesis}, was capable of forming novel materials and nanostructures in nanometric sizes \cite{hirsch2010era}.In its pristine form, graphene acts as a zero-gap semiconductor, imposing constraints on its applicability in certain electronic contexts, including digital transistors. A potential solution to this challenge involves deliberately introducing a gap while safeguarding graphene’s intrinsic electronic characteristics \cite{novoselov2004electric,PhysRevB.82.073403}. Also, an innovative proposal is the search for new allotropic forms derived from carbon. Carbon is the most important chemical element and the theoretical study of its new allotropes can be of great interest \cite{lehmann2007handful}. It is in this vision that research that seeks new allotropic forms of carbon has received outstanding recognition in the scientific community in the sustained development of nanotechnology. The production of substantial quantities of fullerenes C60 and C70, along with their widespread availability, has spurred significant research endeavors in novel carbon nanostructures and materials \cite{kratschmer1990solid,hebard1993buckminsterfullerene}.

The remarkable symmetry of the 12 pentagonal and 20 hexagonal faces, arranged symmetrically in a structure reminiscent of a soccer ball, captures special attention and fuels extensive research in material science, particularly in the field of nanotechnology. Notably, C60 and C70 play crucial roles in materials science  \cite{Yadav2008}.
The initial evidence of the unique nature of these molecules emerged from experiments conducted by Kroto et al., \cite{kroto1985c60} in which laser desorption was employed to generate carbon clusters for mass analysis. Conditions were meticulously optimized to enhance the production of C60 relative to the previously observed even-numbered carbon clusters by Rohfting et al., \cite{rohlfing1988optical} in carbon vapor. Among various identification techniques, mass spectrometry, 13C-NMR and FT-IR has been pivotal in elucidating the fullerene structures
\cite{baena2002fullerenes,benn2011detection}. The occurrence of intensity-optimized mass peaks centered around 60 carbon atoms led to the hypothesis that other fullerene structures, including C70, C76, …, and C86, also exist \cite{mcelvany1992characterization,farre2010first}. 
Fullerenes have garnered significant attention across various scientific disciplines since their discovery. Their distinctive properties and unique physical characteristics set them apart as exceptional and highly valued materials. Efforts have been dedicated to formulate new allotropic forms of carbon \cite{king1996chemical}, However, the pursuit of innovative carbon structures continued. For instance, Terrones et al., computationally investigate the stability and electronic properties of intriguing planar and tubular carbon allotropes \cite{terrones2000new}. They have also discussed the prospective of allotropic forms of carbon \cite{zhang2016carbon,bianco2018carbon,bianco2020carbon}.

Fullerenes find diverse applications, including surface coatings, conductive devices, and the formation of molecular networks \cite{Yadav2008}. A molecular crystal of fullerene behaves as a semiconductor with an approximate bandgap of 1.5 eV, exhibiting characteristics closely resembling those of other semiconductors \cite{morii1997electronic}. In the realm of electronics, it serves conventional roles in applications such as diodes, transistors, photocells, and more. The optical and electronic attributes of fullerenes are primarily contingent on the density of states (DOS) distribution within the mid-gap region of these structures \cite{sachdeva2020optical,shinar1999optical}. Thermal conductivity is one of the parameters to contemplate when applying fullerenes, for example in the production of nanofluids (NFs) \cite{reding2022thermal}.
The smallest fullerene is the C20 dodecahedron, featuring twelve flat faces. Its most symmetrical form is referred to as the regular dodecahedron, recognized for its uniform pentagonal faces.
The regular dodecahedron is the fourth Platonic solid, characterized by its unique structure. The symbol of the universe in ancient Greece \cite{loyd2012old,luminet2003dodecahedral}. The regular dodecahedron, known as dodecahedrane, was initially synthesized in 1982 by Paquette et al., through a 23-step synthesis process. Subsequent to this achievement, alternative synthesis routes have been developed \cite{ternansky1982dodecahedrane,paquette1988chemical}. Dodecahedrane is a stable compound with a melting point of 430°C \cite{lindberg2021strategies}. This compound can be geometrically extended into three dimensions to create a poly-dodecahedrane structure, with covalently attached dodecahedrane units, potentially giving rise to a new carbon allotrope. The design of carbon allotropes entails the deliberate manipulation of carbon atom bonding and coordination, resulting in structures with desirable properties. Various methodologies, including the application of mathematical concepts, have been suggested to generate novel carbon structures with specific properties  \cite{mushtaq2018psl,udvardi2023topological}.
Perhaps the first theoretical carbon allotrope has been proposed by Riley  \cite{gibson1946amorphous}. Since then, hypothetical carbon allotropes have been suggested, employing diverse hybridization  \cite{tong2021orthorhombic,ghorbanali2022two,wei2020new}.  Nonetheless, carbon allotropes characterized by high stability and a sp3 hybridization have garnered relatively limited research attention. A carbon allotrope with a sp3 hybridization, akin to diamond but featuring a new atomic arrangement, is anticipated to possess unique properties. These distinct characteristics render it highly appealing for a range of applications, including superhard materials, optoelectronics, and other areas that demand enhanced functionality  \cite{ghorbanali2023stability,görne2019covalent,ye2023zero,paquette1982dodecahedrane}. The lattice structure of diamond crystals is composed of tetrahedral units, which, owing to their cyclohexane-based configuration, exhibit a near-methane hybridization  \cite{luo2022triatomic,koji2005diamond}. 

Conversely, poly-dodecahedrane is formed through the covalent bonding of cyclopentane rings, resulting in a structure with sp3 hybridization. However, it's worth noting that poly-dodecahedrane comprises cyclopentane units, which are expected to introduce additional angle tension and steric repulsion into the structure  \cite{athreya2022platonic}. 
Interestingly, theoretical studies have revealed that the regular structure of dodecahedrane oligomers preserves structural symmetry during computational optimization. This stability in the oligomeric structures may be attributed to their unique symmetric structural pattern. 
DFT calculations were utilized, followed by vibration frequency computations, to evaluate the electronic density of states ($DOS$), the energy gap between the highest occupied molecular orbital ($E_{HOMO}$) and the lowest unoccupied molecular orbital ($E_{LUMO}$), cohesive energy ($Eg$), and $\Delta G$ for the designed structures. These analyses aimed to explore the stability and electronic characteristics of poly-dodecahedrane \cite{mirderikvand2023novel,delodovici2018protomene}. Up to this point, the domain of theoretically designed carbon allotropes primarily includes structures with unsaturated bonds, in addition to a limited range of other allotropes characterized by saturated structures with a twisted conformation akin to diamond. 
Remarkably, this compound exhibits an exceptionally regular structure, setting it apart from other allotropes, thanks to its unique saturated pentagonal configuration, which remains structurally consistent even during the process of oligomerization  \cite{liu2017new,zhu2020new,fan2019nanoribbons}. Taking into account these features, which encompass its stability, structural consistency, and electronic properties, poly-dodecahedrane can be regarded as a significant carbon allotrope.
\section{Computational Methodology}
\label{CM}

In the computational details, Gaussian 09 program was utilized \cite{gaussian09} for all calculations. The optimal geometry of the designed structures underwent systematic investigation using a DFT framework  \cite{hohenberg1964inhomogeneous,kohn1965self} with $B3LYP$ hybrid functionals, and the 6-311G (d, p) basis set. Additionally, vibration frequencies were computed at the same theoretical level.  The results of structural optimization informed the calculation of electronic density of states ($DOS$) using $GaussSum03$  \cite{cclib}. Based on the DOS outcomes, the energy gap ($Eg$) was determined (Eq.\ref{Eq:01}).

\begin{eqnarray}
    E_{g} = E_{LUMO} - E_{HOMO}
\label{Eq:01}
\end{eqnarray}
Here, $E_{HOMO}$ represents the energy of the highest occupied molecular orbital ($HOMO$) and, $E_{LUMO}$ represents the energy of the lowest unoccupied molecular orbital ($LUMO$). Notably, as the number of dodecahedrane units increases, the Eg  also increases, signifying enhanced stability in the covalent 3D structure of poly-dodecahedrane. $CAM-B3LYP$ method along with the $6-311++G(2df,2pd)$ basis set were employed for calculating the energy of the designed molecules. Furthermore, cohesive energy for all molecules was derived from these energy values. Cohesive energy quantifies the energy required to disassemble a solid into its constituent atoms, transforming them into a collection of neutral free atoms. This value is determined by Eq.\ref{Eq:02}.

\begin{eqnarray}
    E_{coh} = \frac{\left [ E_{tot} - \left ( \sum_{i=1}^{n} n_{i} E_{i} \right )\right ]}{n},
\label{Eq:02}
\end{eqnarray}
where $E_{tot}$, $E_{i}$, and $n_{i}$ represent the total energy of all designed molecules, the atomic energy, and the number of atoms, respectively, is the total number of all atoms.

\section{Results and Discussions}

\subsection{Nanostructural optimization}

\begin{figure}[ht!]
\begin{center}
\includegraphics[angle=0,scale=0.35]{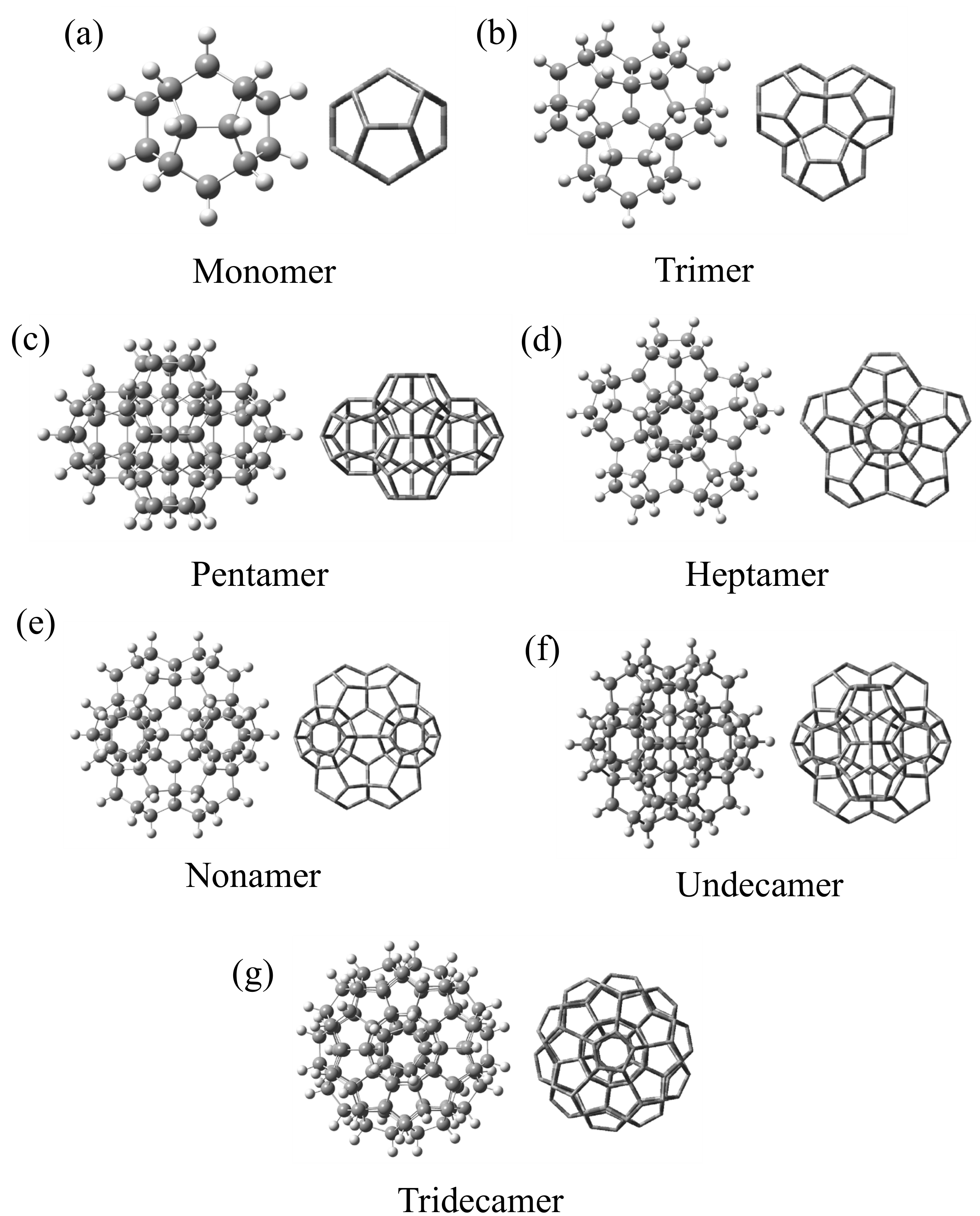}
\caption{\footnotesize{The optimized structures of dodecahedrane oligomers.}}
\label{FIG:01}
\end{center}
\end{figure}

Symmetric forms of oligomers of dodecahedrons $(n=1, 3, 5, 7, 9, 11, 13)$ with the maximum number of quaternary carbons and the minimal carbon-hydrogen bonds necessary for orientation in three dimensions were conceptualized. Subsequently, these structures underwent optimization. As shown in Fig.\ref{FIG:01}. the iner-structure carbons of designed structures are quaternary carbons and all the surface carbons are tertiary carbons.

Structures are optimized to find the stability and minimum energy of systems. Table \ref{TAB:01} presents the physical and chemical parameters of a $13$-unit polydodecahedron, which stands as the largest optimized molecule. To attain the overall minimum energy, it is imperative that the calculated IR spectrum exhibits an absence of negative frequencies. Therefore, the smallest calculated frequencies of the oligomers were carefully examined. As depicted in Table \ref{TAB:02}, the smallest calculated frequencies for the regular dodecahedrane and other optimized structures are all positive.

\begin{table}[hbt!]
    \centering
    \caption{Chemical/physical properties of the $13$ units poly-dodecahedrane.}
    \begin{tabular}{|c|c|}
 \hline
 Parameters  & Value  \\ [0.5ex] 
 \hline\hline
 Intrapentagon $C-C$ distance (for single unit \AA) & $1.56$  \\ 
 \hline
 Cage diameter (for single unit \AA) & $4.37$  \\
 \hline
 Electron affinity $(eV)$ & $0.27$  \\
 \hline
 Ionization potential $(eV)$ & $-6.05$ \\
 \hline
 Binding energy/atom $(eV)$ & $6.69$ \\ 
 \hline
Polarizability ($\textit\AA^{3}$) & $485.76$ \\
 \hline
\end{tabular}
        \label{TAB:01}
\end{table}

\begin{table}[hbt!]
    \centering
    \caption{The calculated smallest frequencies for optimized nanostructures.}
    \begin{tabular}{|c|c|c|c|c|c|c|c|}
 \hline
 Dodecahedrane  & Monomer & Trimer & Pentamer& Heptamr & Nonamer & Undecamer & Tridecamer  \\
 \hline \hline
 Frequency ($cm^{-1}$) & $488.50$ & $243.32$ & $234.24$ & $235.69$ & $172.70$ & $218.34$ & $265.19$ \\ 
 \hline
\end{tabular}
        \label{TAB:02}
\end{table}

$CAM-B3LYPs$ with basis sets $6-311++G(2df,2pd)$ were used to calculate the cohesive energy ($E_{coh}$) of all structures. The cohesive energies calculated serve as a measure of the average bonding strength between atoms within each molecule, and these values are presented in Table \ref{TAB:03}.

\begin{table}[hbt!]
    \centering
    \caption{Calculated values of cohesive energy per atoms for the dodecahedrane oligomers.}
    \begin{tabular}{|c|c|c|c|}
 \hline
 Dodecahedrane oligomer  & $E_{coh}$
$(kcal/mol)$ & Dodecahedrane oligomer & $E_{coh}$
$(kcal/mol)$  \\
 \hline \hline
 Monomer & $135.67$ & Nonamer & $155.30$  \\ 
 \hline
  Trimer & $145.19$ & Undecamer & $158.30$  \\ 
 \hline
  Pentamer & $150.43$ & Tridecamer & $160.42$  \\ 
 \hline
  Heptamr & $153.87$ & --- & ---  \\ 
 \hline
\end{tabular}
        \label{TAB:03}
\end{table}

\begin{figure}[ht!]
\begin{center}
\includegraphics[angle=0,scale=0.35]{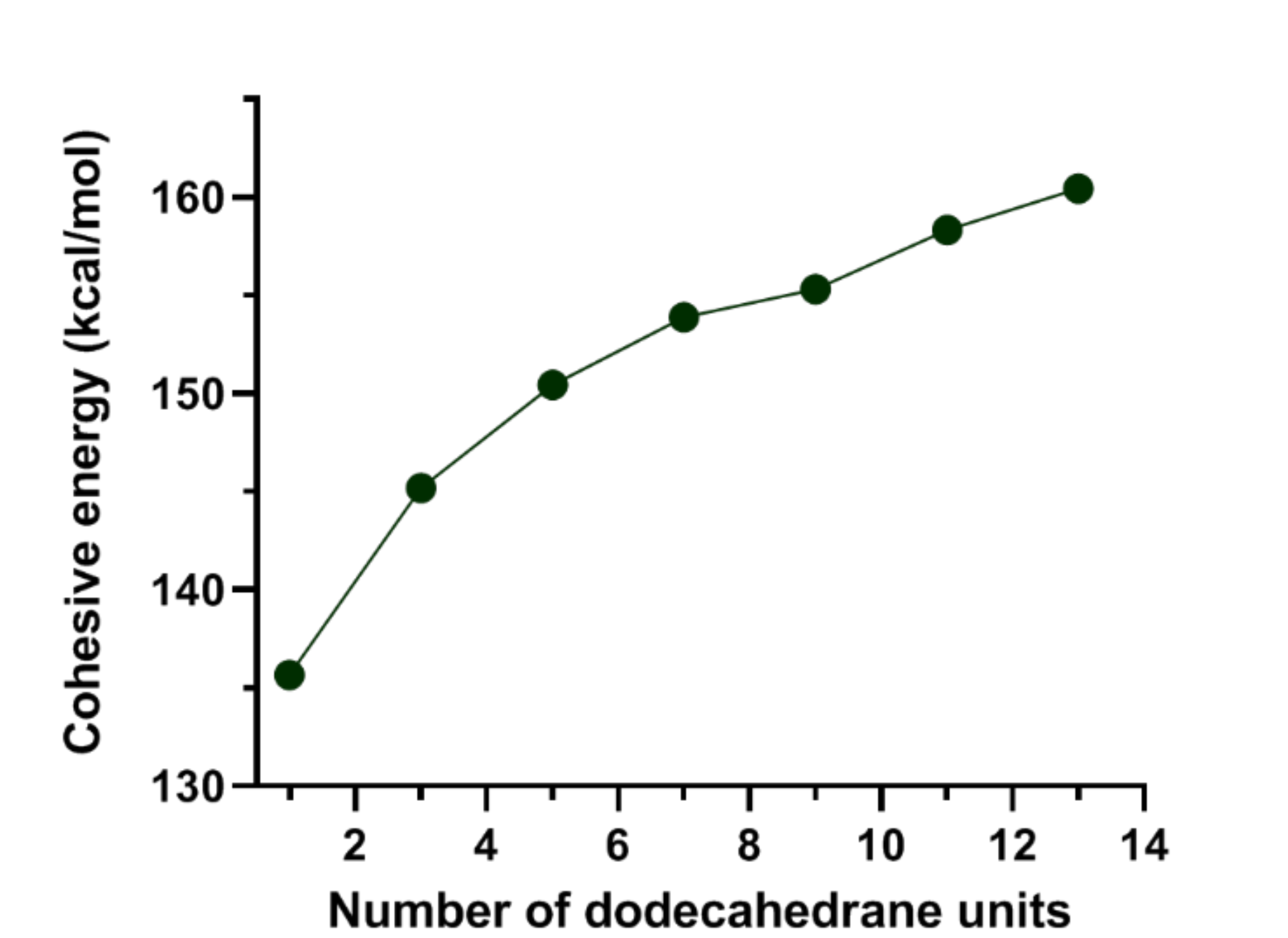}
\caption{\footnotesize{The variation of the cohesive energy with increase the number of dodecahedrane units.}}
\label{FIG:02}
\end{center}
\end{figure}

\begin{figure}[ht!]
\begin{center}
\includegraphics[angle=0,scale=0.35]{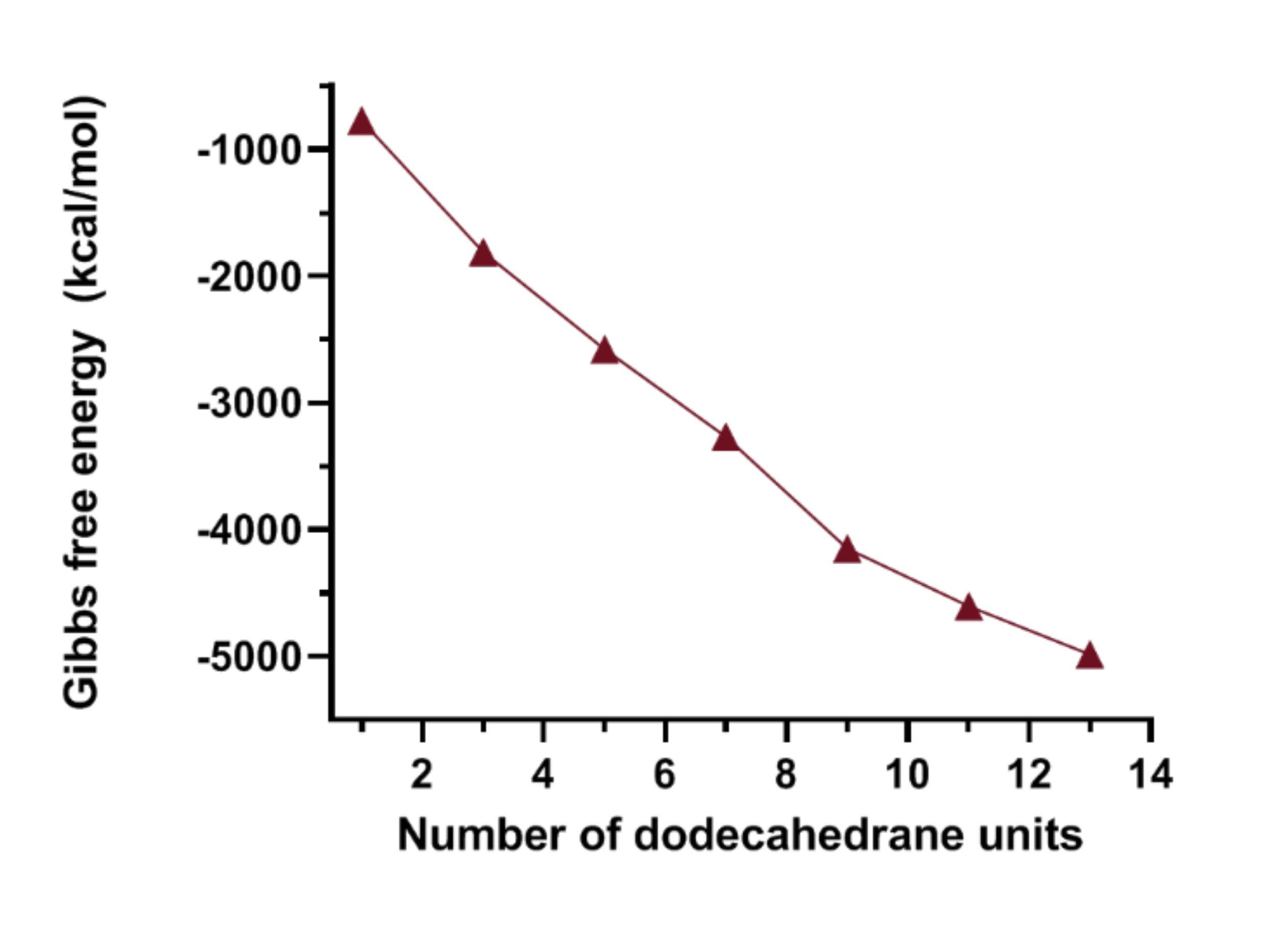}
\caption{\footnotesize{The change of Gibbs free energy as a function of the number of dodecahedrane units.}}
\label{FIG:03}
\end{center}
\end{figure}

As shown in table \ref{TAB:03}, there is a noticeable increase in cohesive energy with the rising number of dodecahedranes. This trend suggests that dodecahedranes tend to extend into three dimensions, resulting in the formation of covalent poly-dodecahedrane solids. The increment in cohesive energy ($E_{coh}$) concerning the number of dodecahedrane units is graphically depicted in Fig. \ref{FIG:02}. Clearly, as the number of dodecahedranes increases, Ecoh exhibits a corresponding increase, indicating that the formation of poly-dodecahedrane is favored due to the propensity of the oligomer units to bond with each other.
Fig. \ref{FIG:03}. shows the Gibbs free energy curve of dodecahedrane oligomers. With an increase in the number of dodecahedrane units, there is a corresponding decrease in the Gibbs free energy, a trend that aligns with the changes observed in the $E_{coh}$ and $Eg$ graphs. This decrease in Gibbs free energy signifies that the formation of poly-dodecahedrane is thermodynamically favorable.

\subsection{Electronic properties}

\begin{figure}[ht!]
\begin{center}
\includegraphics[angle=0,scale=0.3]{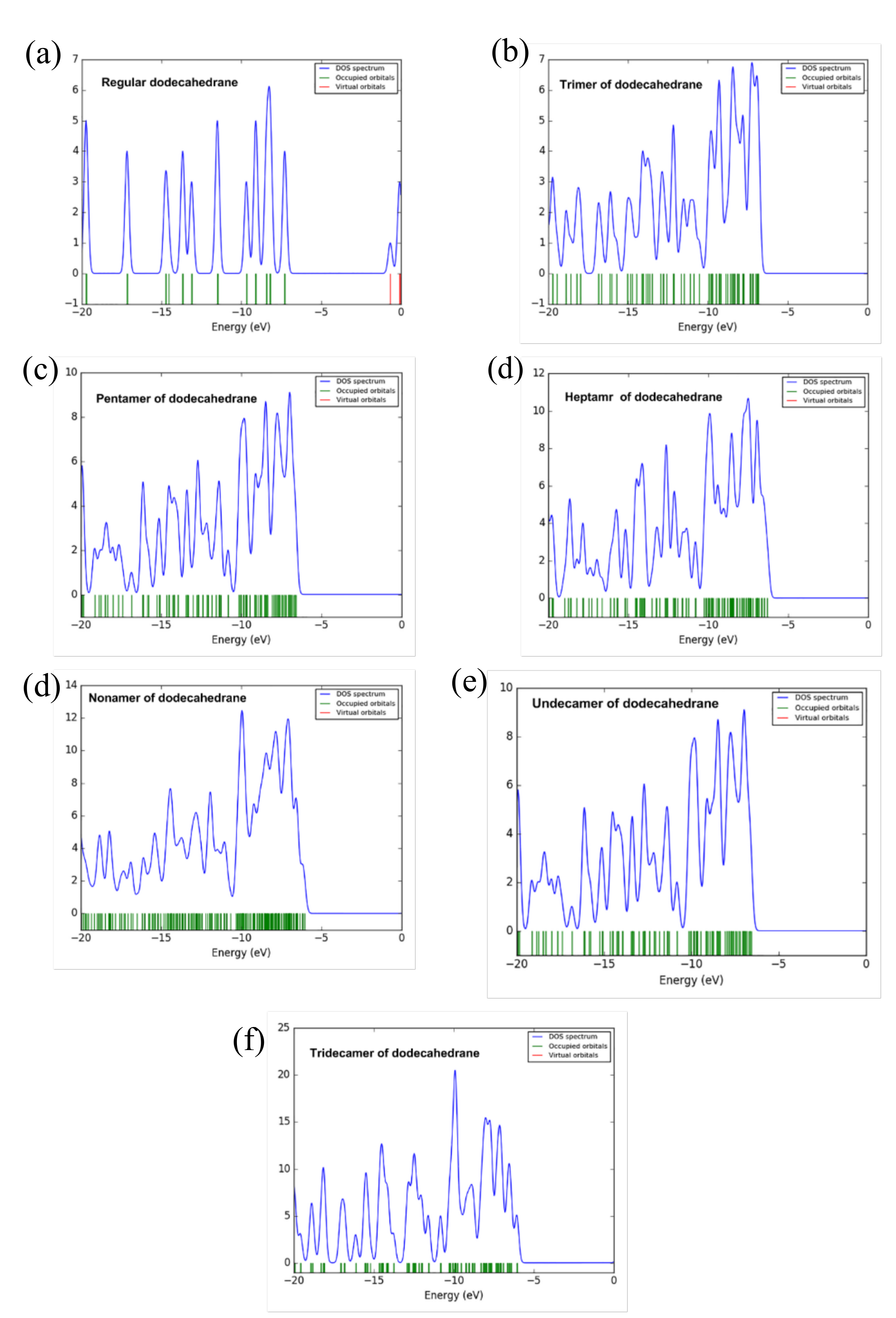}
\caption{\footnotesize{$DOS$ spectra of dodecahedrane oligomers.}}
\label{FIG:04}
\end{center}
\end{figure}

The DOS spectra for the designed molecules are illustrated in Fig. \ref{FIG:04}. Notably, as the energy level intensity rises, the density of electron states within the DOS spectrum also increases. Importantly, the forbidden region, situated between the $E_{HOMO}$ and $E_{LUMO}$, contains no electronic states. The $HOMO$ and $LUMO$ energies were determined based on information extracted from the $DOS$ spectra.

\begin{figure}[ht!]
\begin{center}
\includegraphics[angle=0,scale=0.35]{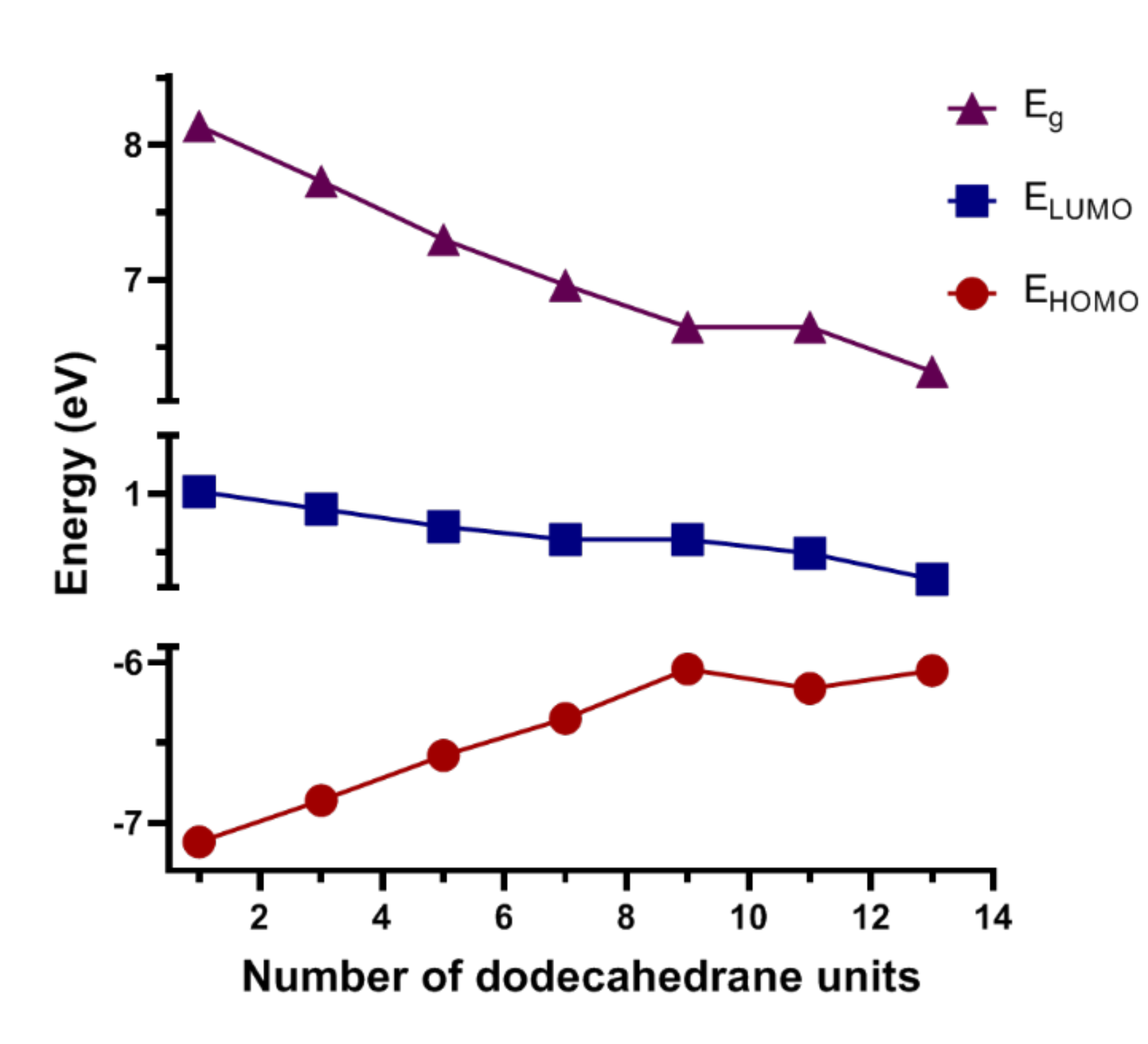}
\caption{\footnotesize{The values of Eg as a function of the number of dodecahedrane units.}}
\label{FIG:05}
\end{center}
\end{figure}

To calculate the $HOMO$-$LUMO$ gap ($Eg$) for these structures, we computed the energies of the HOMO and LUMO, which are documented in \ref{TAB:04}. Interestingly, as the number of dodecahedrane units increases, the HOMO energy rises while the LUMO energy decreases, resulting in a reduction of the $Eg$ value. Fig. \ref{FIG:05}, visually demonstrates that with an increasing number of dodecahedral units, both the $HOMO$ and $LUMO$ energy diagrams converge, leading to a noticeable decrease in the $Eg$ value.

\begin{table}[hbt!]
    \centering
    \caption{The energies of HOMO, LUMO and Eg for these nanostructures.}
    \begin{tabular}{|c|c|c|c|}
 \hline
 Dodecahedrane oligomer  & $E_{HOMO}(eV)$ & $E_{LUMO}(eV)$ & $Eg(eV)$ \\
 \hline \hline
 Monomer & $-7.12$ & $1.02$ & $8.14$  \\ 
 \hline
  Trimer & $-6.86$ & $0.87$ & $7.73$  \\ 
 \hline
  Pentamer & $-6.58$ & $0.72$  & $7.30$  \\ 
 \hline
  Heptamr & $-6.35$ & $0.61$ & $6.96$  \\ 
 \hline
 Nonamer & $-6.04$ & $0.61$ & $6.65$  \\ 
 \hline
 Undecamer & $-6.26$ & $0.51$ & $6.77$  \\
  \hline
  Tridecamer & $-6.05$ & $0.27$ & $6.32$  \\
     \hline
\end{tabular}
        \label{TAB:04}
\end{table}

\begin{figure}[ht!]
\begin{center}
\includegraphics[angle=0,scale=0.6]{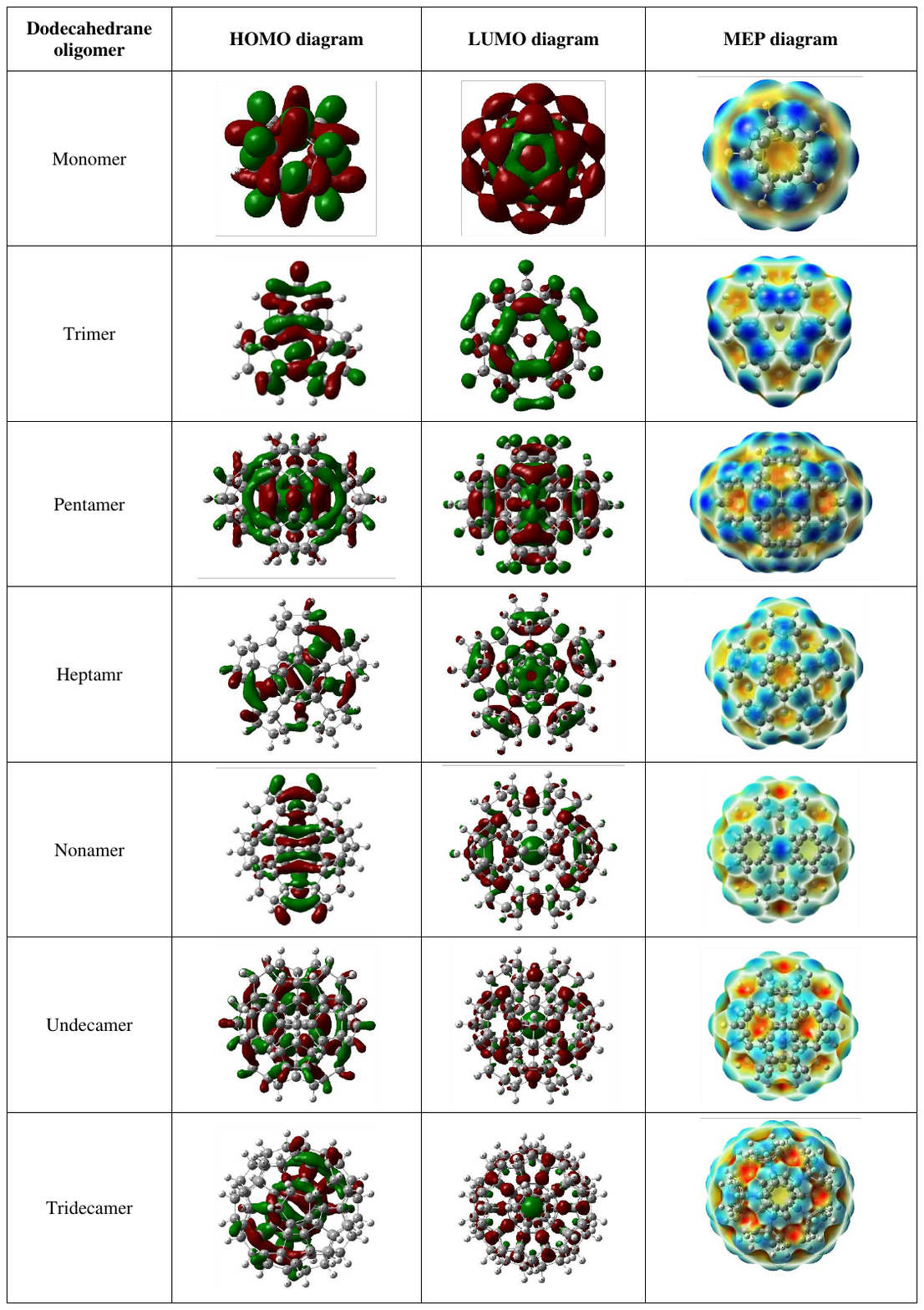}
\caption{\footnotesize{The diagrams of HOMO, LUMO and molecular electrostatic potential (MEP) of dodecahedrane oligomers.}}
\label{FIG:06}
\end{center}
\end{figure}

The diagrams illustrating the HOMO, LUMO, and molecular electrostatic potential (MEP) of dodecahedrane oligomers are presented in Fig. \ref{FIG:06}. Remarkably, even after structural optimization, the density distribution remains symmetric, and the dodecahedrane oligomer units maintain their regular configuration. The cyclopentane rings exhibit resistance to twisting in order to minimize energy, opting instead to maintain a completely flat configuration within the structure. As a result, the overall nanostructural order is preserved, presenting itself as a combination of symmetric dodecahedrons. Fascinatingly, all oligomers displayed nearly identical angles falling within a narrow range of $108^{\circ}.5$ to $110^{\circ}$. This angle range closely approximates the ideal angle for methane ($109^{\circ}.28$), which represents the minimum energy and maximum stability. Significantly, this value surpasses other unsaturated and saturated carbon allotropes, rendering it a superior property.

An $NBO$ analysis was performed on the tridecamer molecule using the $DFT/B3LYP/6-311G (d,p)$ level to elucidate the hybridization of carbon atoms at different positions, as illustrated in Fig. \ref{FIG:06}. The analysis uncovered that the innermost carbons at positions 85 and 86 exhibit $sp^{3.03}$ hybridization, coinciding with the vertices of the central dodecahedron. Meanwhile, the intermediate carbons at positions 93 and 108 display $sp^{3.16}$ and $sp^{3.15}$ hybridization, respectively. The carbon atom located on the surface at position 106 demonstrates $sp^{3}$.15 hybridization. It's worth highlighting that the bond lengths exhibit variation among these carbon atoms. Upon reflecting on the findings, it becomes evident that if the network is extended, all units will eventually acquire the characteristics of the innermost unit, whose hybridization resembles that of diamond. This suggests that expanding the network will enhance the system's stability. Additional information regarding the distances and hybridizations can be found in table  \ref{TAB:05}.

\begin{figure}[ht!]
\begin{center}
\includegraphics[angle=0,scale=0.50]{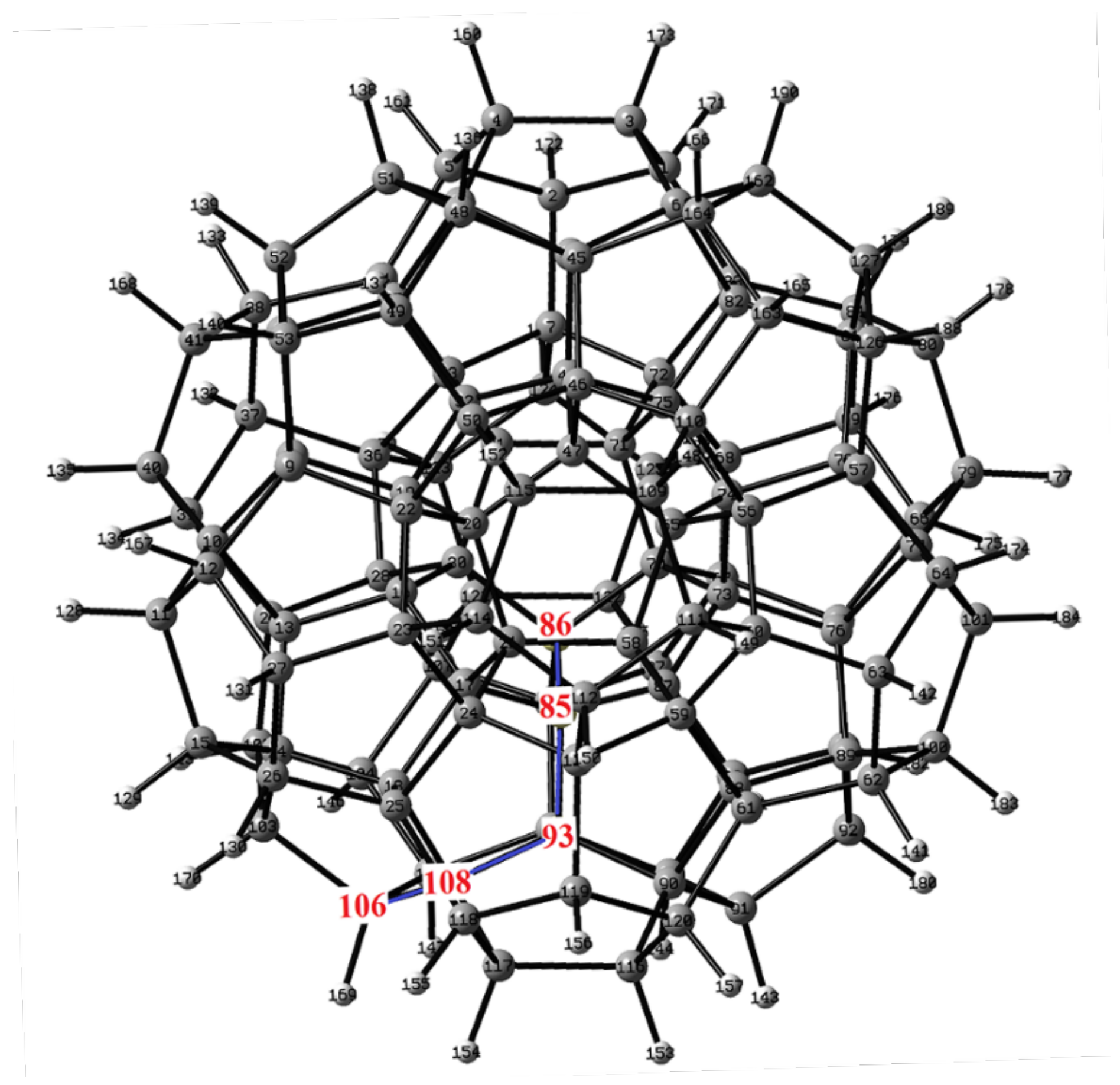}
\caption{\footnotesize{Diagram the positions of the outer and internal atoms within the tridecahedron oligomer.}}
\label{FIG:07}
\end{center}
\end{figure}

\begin{table}[hbt!]
    \centering
    \caption{Hybridization in NBO basis and C-C distances within the tridecahedron oligomer}
    \begin{tabular}{|c|c|c|c|}
 \hline
 Atom number  & Coefficients/ Bond orbital & Hybridization & distance ($\text{\AA}$)  \\
 \hline \hline
 $C_{85}$ & $0.7071/s( 24.84\%)p( 75.16\%)$ & $sp^{3.03}$ & $C_{85}-C_{86} = 1.48$  \\ 
 \hline
  $C_{86}$ & $0.7071/s( 24.83\%)p( 75.17\%)$ & $sp^{3.03}$ & $C_{85} - C_{86} = 1.48$  \\ 
 \hline
  $C_{93}$ & $0.7014/s( 24.05\%)p( 75.95\%)$ & $sp^{3.16}$ & $C_{85} - C_{93} = 1.49$  \\ 
 \hline
  $C_{108}$ & $0.7036/s( 24.08\%)p( 75.92\%)$ & $sp^{3.15}$ & $C_{93} - C_{108} = 1.57$  \\ 
 \hline
  $C_{106}$ & $0.7002/s( 25.59\%)p( 74.41\%)$ & $sp^{2.91}$ & $C_{106} - C_{1.08} = 1.59$   \\ 
 \hline

\end{tabular}
        \label{TAB:05}
\end{table}

The total dipole moment of the 13-unit structure amounts to zero $(X= 0.0000, Y= 0.0000, Z= 0.0000)$. As anticipated, given the absence of free electrons, this structure exhibits minimal optical activity. These characteristics, in conjunction with other properties, bear a striking resemblance to the structure of diamond. Anticipated characteristics of this composition include very low electrical conductivity and very high thermal conductivity. Additionally, given its structural similarity to diamond, it is likely to possess a similar density and exhibit exceptionally high hardness. From a chemical perspective, it is conceivable to suggest a synthesis method for this compound, where the structural evolution leads to the formation of interconnected pentagonal networks. One potential method involves the coupling of two monomers, namely full-chloride ethane and propane, in the presence of an coupling catalyst metal (alkali metals) within an oxygen and moisture-free atmosphere. The coupling of hexachloroethane and octachloropropane may force the system to generate cyclopentane units and establish connections among these rings in three dimensions.

\newpage
\section{Conclusions and remarks}

In this investigation, a series of systematically designed dodecahedrane oligomers, each comprising varying constituent units, underwent precise examination of their structural properties and stability. The calculations revealed that the cohesive energies of these oligomers increased with the number of dodecahedrane units. Additionally, as the number of dodecahedrane units rose, the gap between the $HOMO$ and $LUMO$ energies, along with the Gibbs free energy values, decreased.

Additionally, a natural bond orbital ($NBO$) analysis was conducted on the $13$-units oligomer, revealing that various carbon atoms exhibited near-$sp^{3}$ hybridization. Overall, the calculated results strongly suggest that the stability of the oligomers increases with the greater number of dodecahedrane units. These findings collectively imply that poly-dodecahedrane is theoretically stable and holds the potential to be recognized as a novel carbon allotrope.

\newpage
\section{Acknowledgements}

S.H. and E.N. acknowledges Department of Organic Chemistry, Faculty of Chemistry, Lorestan University, Khorramabad, Lorestan, Iran . J.M.S. acknowledges CAPES, CNPq, FAPESP, FAPEPI and Instituto Federal de Educação, Ciência e Tecnologia do Piauí -- IFPI. 
\newpage
\bibliographystyle{elsarticle-num}
\bibliography{biblio.bib}
\end{document}